\documentclass[structabstract]{aa}  
\usepackage{graphicx}
\usepackage{txfonts}
\usepackage{lscape,longtable}
\usepackage{natbib}
\bibpunct{(}{)}{;}{a}{}{,}
\begin{document}
   \title{Beyond the pseudo-time-dependent approach: chemical models of dense core precursors}

%\subtitle{I. Overviewing the $\kappa$-mechanism}

   \author{G. E. Hassel
          \inst{1}
          \and
          E. Herbst \inst{1,2}
          \and
          E. A. Bergin \inst{3}
          }

   \institute{Department of Physics, The Ohio State University,
             Columbus, OH 43210, USA\\
              \email{ghassel@mps.ohio-state.edu}
         \and
             Departments of Astronomy and Chemistry, The Ohio State University,
                Columbus, OH 43210, USA\\
          \and
             Department of Astronomy, University of Michigan, 825 Dennison Building, Ann Arbor, MI 48109, USA\\
           }

   \date{}

% \abstract{}{}{}{}{} 
% 5 {} token are mandatory

  \abstract
  {Chemical models of dense cloud cores often utilize the so-called pseudo-time-dependent approximation, in which the physical conditions are held fixed and uniform as the chemistry occurs.  
In this approximation, the initial abundances chosen, which are totally atomic in nature except for molecular hydrogen, are artificial.   A more detailed approach to the chemistry of dense cold cores should include %what happens 
the physical evolution during their early stages of formation.}
   {Our major goal is to investigate the initial synthesis of molecular ices and gas-phase molecules  as cold molecular gas begins to form behind a shock in the diffuse interstellar medium.  The abundances calculated as the conditions evolve can then be utilized as reasonable initial conditions for a theory of the chemistry of dense cores.}
   {Hydrodynamic shock-wave simulations of the early stages of cold core formation are used to determine the time-dependent physical conditions for a gas-grain chemical network.   We follow the cold post-shock molecular evolution of ices and gas-phase molecules for a range of visual extinction up to $A_{\rm V} \approx 3$, which increases with time. At higher extinction, self-gravity becomes important.  }
   { As the newly condensed gas enters its cool post-shock phase, a large amount of CO is produced in the gas.  As the CO forms,  water ice is produced on grains, while accretion of CO produces CO ice.  The production of CO$_{2}$ ice from CO occurs via several surface mechanisms, while the production of CH$_4$ ice is slowed by gas-phase conversion of C into CO. }
   %Column densities of major ice and gas-phase species are tabulated at various values of $A_V$. }
   {}

               
 \titlerunning{Chemical models of dense core precursors}
%Chemical models of a dense core in formation}
     \authorrunning{Hassel et al.}

   \maketitle
%
%________________________________________________________________

\section{Introduction}

   The formation of molecular clouds from the diffuse atomic interstellar medium  has been the subject of much interest. Proposed mechanisms have been reviewed in the recent literature \citep{ABI09, BT07, MO07}.  Despite this interest, most chemical models of cold dense interstellar clouds utilize the simple pseudo-time-dependent approximation, in which the physical conditions are homogeneous and fixed, beginning from an already dense, cold, and darkened state with all H in H$_2$.  Moreover, the initial abundances for heavy elements are assumed to be atomic. Yet, the high abundance of molecular hydrogen in diffuse clouds suggests that molecules can be synthesized as clouds form as well as during their existence.  One approach to dense cloud formation, that of hydrodynamic shock waves, has been studied by \citet{Bergin04} (hereafter B04), who showed that the gas-phase molecules H$_2$ and CO are produced early in the formation of the cloud.  In this paper, we revisit the approach of B04, and include a more complex treatment of the gas-grain chemistry that evolves in tandem with the post-shock physical conditions.  
We focus on the initial production of major solid phase and gaseous molecules in the post-shock material as the extinction gradually increases.   
  
Prior investigations have considered some aspects of the problem we discuss here.   Early dynamic models of dense and diffuse clouds were reviewed by \citet{Wi88}. Cyclic models involving shocks were studied by \citet{NW92}. Our approach is similar to the constant-pressure collapse model of \citet{Pineau91}, although that paper did not include grain surface chemistry.  Such chemistry was included by \citet{RH01III} in a pseudo-time-dependent model of the formation of mantle ices using an earlier version of our gas-grain network.  These authors were mainly interested in explaining the formation of CO$_{2}$ ice, which had been under-produced in their previous models.  A pseudo-time-dependent approach with surface chemistry was also used to study the water ice distribution at various conditions throughout the Taurus dark cloud \citep{Nguyen02}.

%__________________________________________________________________

\section{Hydrodynamical Model}

%__________________________________________________________________ 

%\begin{figure*}[b]
\begin{figure}
\centering
\includegraphics[angle=90,scale=.4]{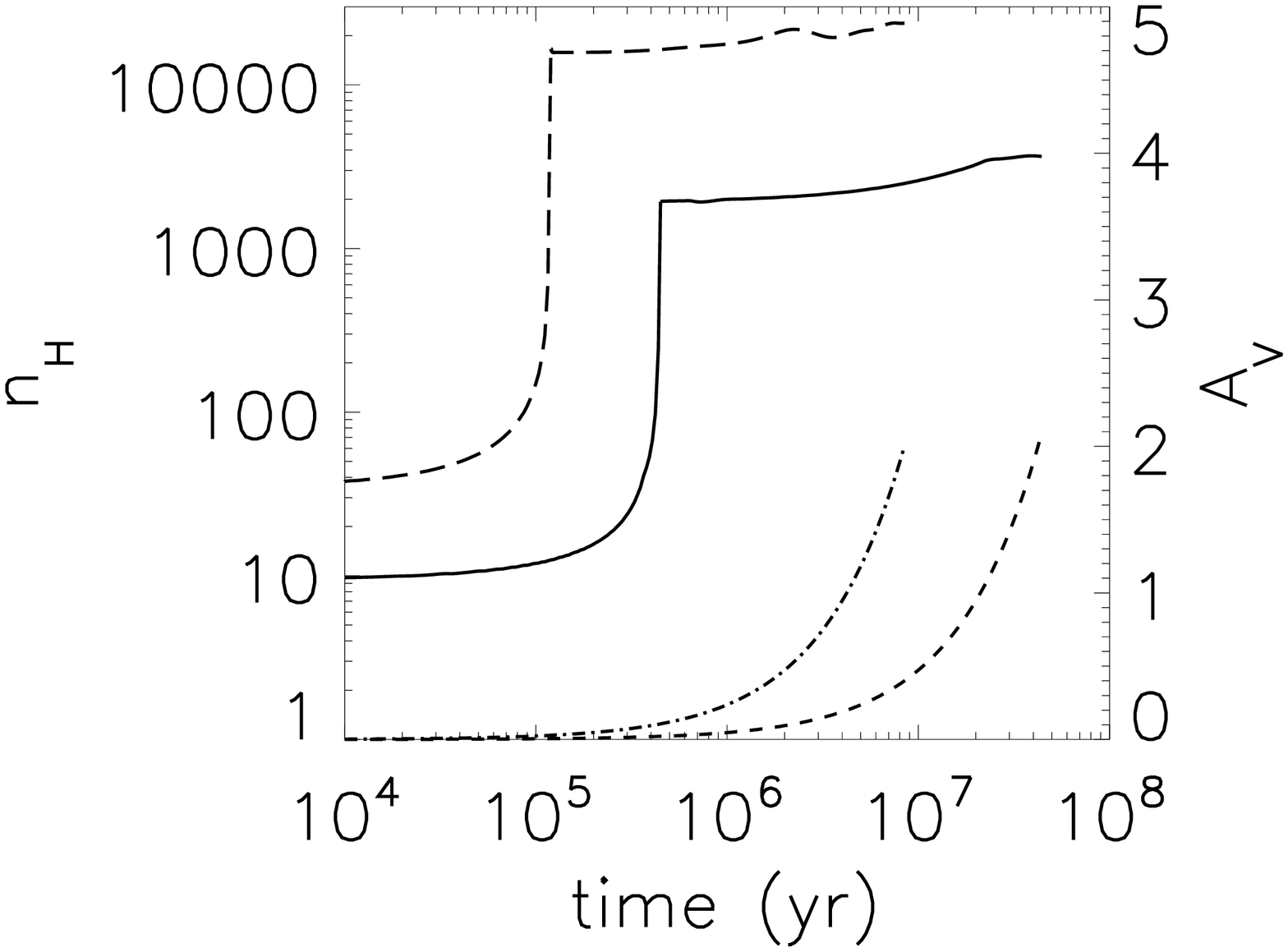}
\includegraphics[angle=90,scale=.4]{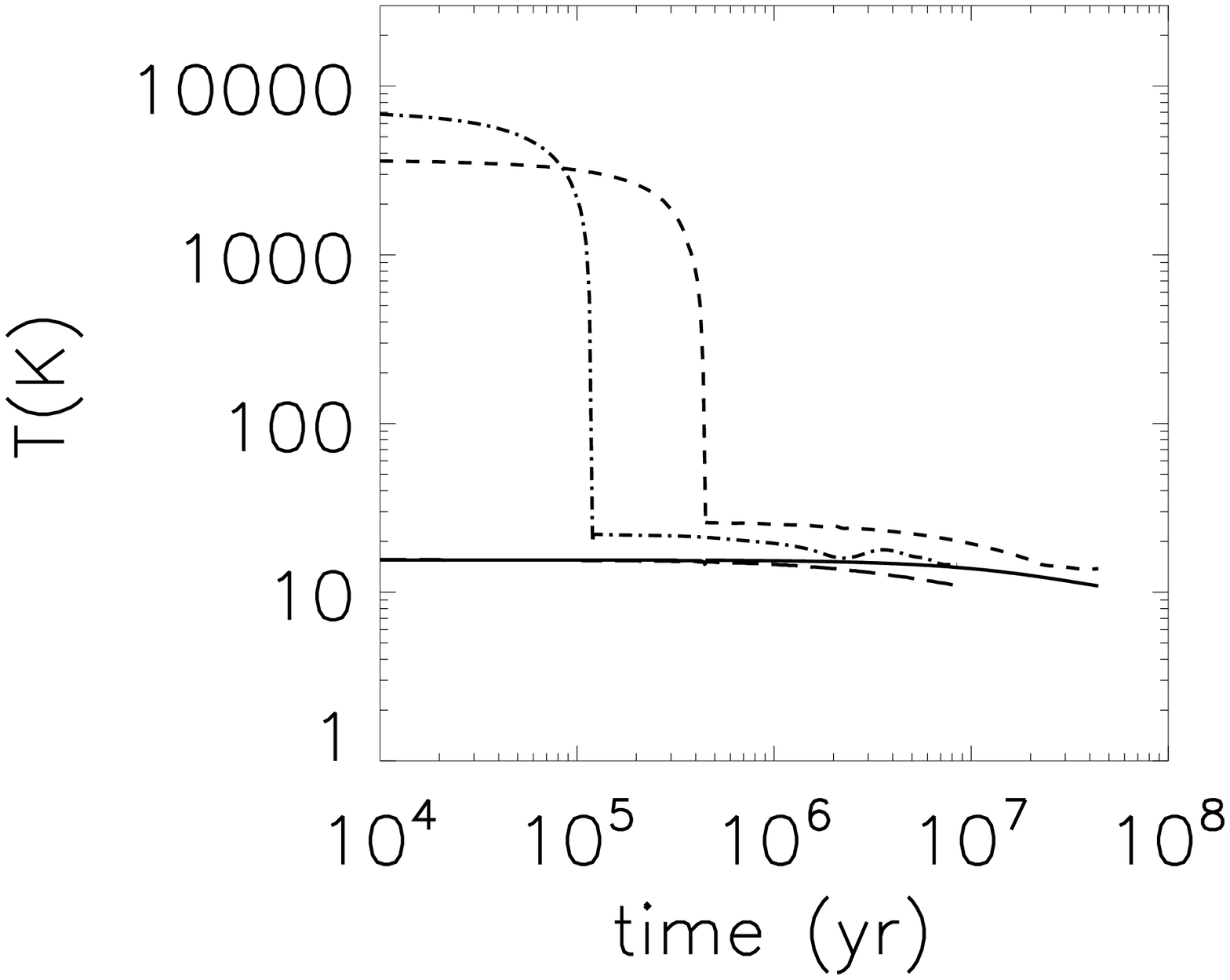}
\caption{Hydrodynamic results for Models 3~\&~4 with dust temperature calculations.  Model 3 parameters are represented by  solid lines (density, dust temperature with $T_{\rm d,0} = 15$ K), and  dashed lines (extinction, gas temperature).  Model 4 parameters are represented by  long dashed lines (density, dust temperature), and dotted-dashed lines (extinction, gas temperature). }
\label{Evolvehydro}
%\end{figure*}
\end{figure}

We have used four shock models based on the results presented in B04; the shock speed, initial total hydrogen density, and fractional abundance of gaseous H$_{2}$ if any are listed in Table \ref{tbl-shockmods}.  Of these four models, the first three were taken directly from B04 whereas the fourth model, starting with the densest gas, was run using unpublished B04 model results, the motivation being to explore  greater initial densities.  

 The shock initially heats the low density gas, which cools via atomic fine structure emission.  At the cooling timescale, the gas-phase temperature drops to the equilibrium of C$^+$ cooling and photoelectric heating, while the density rises.  It is important to note that the chemistry modeled in this work is {\it not} shock chemistry, such as the evolution during the hot phase of hydrodynamic shocks studied by \citet{MitchellWatt} and associated papers.  Rather, all of the relevant chemical evolution occurs in the post-shock phase where the initially atomic gas has become dense ($n_{\rm H} \ge 1000$ cm$^{-3}$) and cold ($T_{\rm g} \lesssim 20$ K), but critically is still at low extinction.  We present the conditions at the onset time of this stage when gas temperature and density reach near-constant conditions and identify it as the ``discontinuity time'', $t_{\rm disc}$, in Table~\ref{tbl-shockmods}.  As time proceeds in our models, the major change in the physical state is increasing dust extinction, which is key for the growth of molecular abundances.  Table~\ref{tbl-shockmods} also contains information concerning the conditions of the post-shock cold and dense gas later when $A_V$ has reached 2.0.

Some physical results for models 3 and 4, which possess the lowest and highest density, respectively, in the post-shock gas, are displayed in Figure~\ref{Evolvehydro} starting at a time of 10$^{4}$ yr after the shock.  Similar plots for models 1 and 2 previously appeared in Figure 2 of B04.  As the gas temperature cools and the density increases, the visual extinction  and column slowly build up over time.  The results of B04 follow the time dependence of $A_V$ until it reaches a value of a few, which can take upwards of 10 Myr, as shown in Figure~\ref{Evolvehydro} and Table~\ref{tbl-shockmods}.  To extend this increase to slightly greater times, one can estimate $A_V$ by equation (4) in B04.  The density and gas temperature are assumed not to change from their final state during this extension.  It is likely that beyond $A_V \sim 2-3$ mag  our simulations are strongly affected by self-gravity \citep{hbb01}.  The same chemical-dynamical treatment cannot therefore be used realistically to reach fully formed cloud cores of higher extinction (e.g., TMC-1).  However,  the model results, especially for the ices, can be compared with observations of cores of low extinction in assemblies such as the Taurus cloud \citep[and references therein]{dougco207}.  In addition, results for CO(g) can be compared with observed values in diffuse and translucent regions.  Perhaps most importantly, the results can also be used as more realistic initial abundances for normal dense core chemistry than the gaseous mainly atomic abundances used in the pseudo-time-dependent treatments.  It is also possible to follow the dynamics and chemistry until cold dense cores form at high extinction with a more detailed hydrodynamical model \citep{Aikawa05}.

%_____________________________________________________________
%
\begin{table*}
\begin{center}
\caption{Shock Models}             
\label{tbl-shockmods}      
\centering                        
\begin{tabular}{l c c c c c c c c c c}        
\hline\hline                 
Model & $v_{\rm shock}$ (km s$^{-1}$) & $n_{\rm H,0}$ (cm$^{-3}$) & $X_{\rm H_2,0}$ & $t_{\rm disc}$$^1$ (yr) & $n_{\rm H,disc}$$^1$ (cm$^{-3}$) & $T_{\rm g,disc}$$^1$ (K) &  $t_{\rm 2}$$^2$ (yr) & $n_{\rm H,2}$$^2$ (cm$^{-3}$) & $T_{\rm g,2}$$^2$ (K) & size$^2$ (pc)\\ 
\hline                        
1 & 20 & 1 & 0    & 1.2(+06)$^3$ & 2.7(+03) & 21.8 &  6.0(+07)    &  5.7(+03) & 10.7 & 0.21\\    
2 & 30 & 1 & 0    & 2.5(+05)     & 6.7(+03) & 20.5 &  4.1(+07)    &  9.5(+03) & 14.3 & 0.13\\    
3 & 10 & 3 & 0.25 & 4.4(+05)     & 1.9(+03) & 25.9 &  4.2(+07)    &  3.7(+03) & 13.7 & 0.52 \\ 
4 & 15 & 10 & 0   & 1.2(+05)     & 1.7(+04) & 20.9 &  8.4(+06)    &  2.4(+04) & 14.6 & 0.05 \\   
\hline                                   
\end{tabular}
\end{center}
$^1$ Refers to values at discontinuity.  See, e.g., Figure~\ref{Evolvehydro}.\\
$^2$ Refers to values at $A_V\approx 2$.\\
$^3$ {a(b) = a $\times 10^{\rm b}$} \\
\end{table*}
%
%_____________________________________________________________

To include surface chemistry during the post-shock cooling,  we determined the evolution of the dust temperature by equating the rate of heating by interstellar radiation, $\Gamma_{\rm UV}$ \citep{cuppen06, Zucconi01, bohhuf83, drainebertoldi96, dustbook03}, and by collisions with gas molecules, $\Gamma_{\rm col}$, with the rate of cooling by thermal emission, $\Lambda_{\rm IR}$ \citep{drainelee84}, and by evaporation $\Lambda_{\rm evap}$.  The computed dust temperature profiles, included in Figure~\ref{Evolvehydro}, begin with $T_{\rm d}\simeq 15$ K regardless of the properties of the gas, and are thus labeled as $T_{\rm d,0}= 15$ K throughout this paper.  However, this temperature falls slightly below the range measured by $COBE$ for the diffuse ISM \citep[$T_{\rm d} \sim 16-23$~K;][]{Reach95}, so we also calculate an increased dust temperature profile with an increased radiation field, starting at $T_{\rm d,0}= 20$ K.  The dust temperature evolution is dominated by the radiative processes within the density range of these simulations, and gradually decreases from the initial $T_{\rm d,0}$ value as $A_V$ increases.  For all the models considered here, the value of $T_{\rm d}$ consistently evolves from 15.5 K at the discontinuity to 11.0 K at $A_V=2$ for the cooler profile and from 20.2 K to 14.4 K for the warmer profile.

%_____________________________________________________________
\section{Chemical Model and Modifications}

We have used the Ohio State (OSU) gas-grain reaction network \citep{hhl92, gpch06, GH06, gwh07}.  This network includes 6323 reactions involving 655 species in an expanded form including two new types of processes.  We adopt a single grain size of $r_{\rm d}=0.1~\mu$m and a Rice-Ramsperger-Kessel (RRK) parameter for reactive desorption of 0.01 \citep{gpch06}.  As in B04, we use a slightly enhanced interstellar radiation field of $G_0 =1.7$, in units of the Habing field \citep{draine78}.   The cosmic ray ionization rate $\zeta_{\rm H}$ is set at $1.3 \times 10^{-17}$ s$^{-1}$.  The elemental abundances used are high in metals;  these consist of the  $\zeta$ Oph abundances used by B04  with the exception of Ne, Ar, Ca, and Ni, which are not in our network.  We have included the elements P, Na, and Cl, which do appear in the OSU networks but not in B04, at their so-called ``high-metal'' values \citep{GH06}.  The term ``high'' is used to distinguish the initial estimate from those used to estimate depletions in initially dense models \citep{gwh07}.   

The self-shielding of H$_2$ is treated in the present models using the formalism of \citet{lee96} as in the previously mentioned networks.  To treat the self shielding of CO(g) in our model at low $A_V$ more properly, we have adopted photodissociation rates based on the results of the Meudon PDR code \citep{Meudon}.  Specifically, we obtain individual photodissociation rates at representative conditions spanning the range of $n_{\rm H}$ and $A_V$ variation for each of the four models, and determine a unique exponential fit to the form $k = \alpha \exp(-\gamma A_{\rm V})$  for each model from those results.  Our shielding results are in reasonable agreement with those of B04, whose H$_{2}$ shielding is based on \cite{drainebertoldi96}.
     %To  treat the self-shielding of CO(g) in our model at low $A_{\rm V}$ more properly, we have adopted photodissociation rates  based on results of the Meudon PDR code \citep{Meudon} within a range of relevant conditions for each model.  
     On the grain surfaces, photodesorption 
     via external UV photons is an important process to include.  We have included photodesorption processes for the four species studied in the laboratory by \citet{ObergCO07,  ObergWater, ObergPHO}: CO(s), N$_{2}$(s), H$_{2}$O(s), and CO$_{2}$(s), where (s) refers to the solid phase.

The surface reactions used in the gas-grain chemical network normally involve the diffusion of one or more reaction partners across surface sites via thermal hopping, a process known as the Langmuir-Hinshelwood mechanism.  At low temperatures, H atoms are a most important surface reactant, and are responsible for the hydrogenation of heavy atoms landing on grains into their saturated forms (e.g. O into H$_{2}$O, C into CH$_{4}$).  Of the major ice species in our model (H$_2$O, CO, CO$_{2}$), CO$_{2}$ has proven to be the most difficult to produce on surfaces via such processes \citep[e.g.][]{RH01III}.  We have adopted an activation energy $E_{\rm A}$ = 290 K, as measured by \citet{RVMP01}, for the important CO$_{2}$(s) formation reaction between CO(s) and O(s) \citep{Nummelin01}.  Following the approach of \citet{RH01III}, we also consider a decreased activation energy $E_{\rm A}$ of 130 K \citep[see, for example,][]{gdh86, FEA79, TH82}. 

Gas-phase species can also react directly with surface species via the Eley-Rideal mechanism \citep{RH01III}.  To be efficient, such a mechanism requires a reactant with reasonably high surface coverage, and CO(s) meets this criterion once CO(g) starts to accrete.  To possibly enhance the rate of CO$_{2}$ formation, the Eley-Rideal reaction $\rm{O(g)}+\rm{CO(s)} \rightarrow \rm{CO_{2}(s)}$ was added to the network, with a rate coefficient $k_{\rm ER}$  given by
\begin{equation}
\label{OCOerrate}
k_{\rm ER} = k_{\rm acc,O} \theta_{\rm CO(s)} \exp\left(-E_{\rm A,E-R}/T_{\rm g}\right),
\end{equation}
where $k_{\rm acc,O}$ is the accretion rate (s$^{-1}$) of O(g) onto the grain surface, $\theta_{\rm CO(s)}$ is the average fractional surface coverage of CO(s), and the Boltzmann factor depends on a reaction barrier, $E_{\rm A, E-R}$.  A value of $E_{\rm A,E-R}= 0$ K is assumed in order to investigate the full capability of this reaction.

To consider a reasonable portion of parameter space, we have run a considerable number of models.  The ones chosen for discussion comprise all four shock models listed in Table~\ref{tbl-shockmods}, each with two different evolutions of granular temperature and two different barriers for the diffusive O $+$ CO reaction.
Models that contain $T_{\rm d,0} = 15$~K and $E_{\rm A}$ = 290 K are labeled the A set, while models that contain $T_{\rm d,0} = 20$~K and $E_{\rm A}$ = 130 K are labeled the B set.  With the shock models 1-4, we label the chemical models 1-A, 1-B, etc.  Models with other sets of varied parameters were also considered, but our choices for A and B parameters span the range of variability of results.

%_____________________________________________________________

\section{Results}
\label{resultssection}

We report our principal results as column densities rather than fractional abundances  
 to facilitate comparison with the available observations (see next section).  The column densities are obtained from the calculated fractional abundances at a given visual extinction by using a relation between the visual extinction and the total column density for hydrogen, $N_{\rm H}=1.9\times10^{21} A_V$ cm$^{-2}$ mag$^{-1}$ \citep{dustbook03}, although there is some variation in the adopted value of this conversion factor in the literature.  

As in B04, the gaseous species H$_{2}$ and CO form after an enhanced density $n_{\rm H,disc}$  at low temperatures even though the extinction is very low.  A variety of other species start to form, both in the ice phase and in the gas.  After H$_{2}$ and CO, the most abundant gas-phase species by $A_V=2$ include N$_{2}$, CN, H$_{2}$O, and CH$_4$, although the last three possess columns of $\approx 10^{14}$ cm$^{-2}$ or less, nearly three orders of magnitude smaller than the first three.  The high-metal elemental abundances maintain a high electron fractional abundance exceeding 10$^{-6}$, which diminishes the power of the ion-molecule chemistry to produce larger molecules. There are no molecular ions among the top 50 species of any model considered at this time, nor does the largest abundance of a molecular ion come within six orders of magnitude of the CO(g) abundance.

\begin{figure*}[b]
\centering
\includegraphics[angle=90,scale=.4]{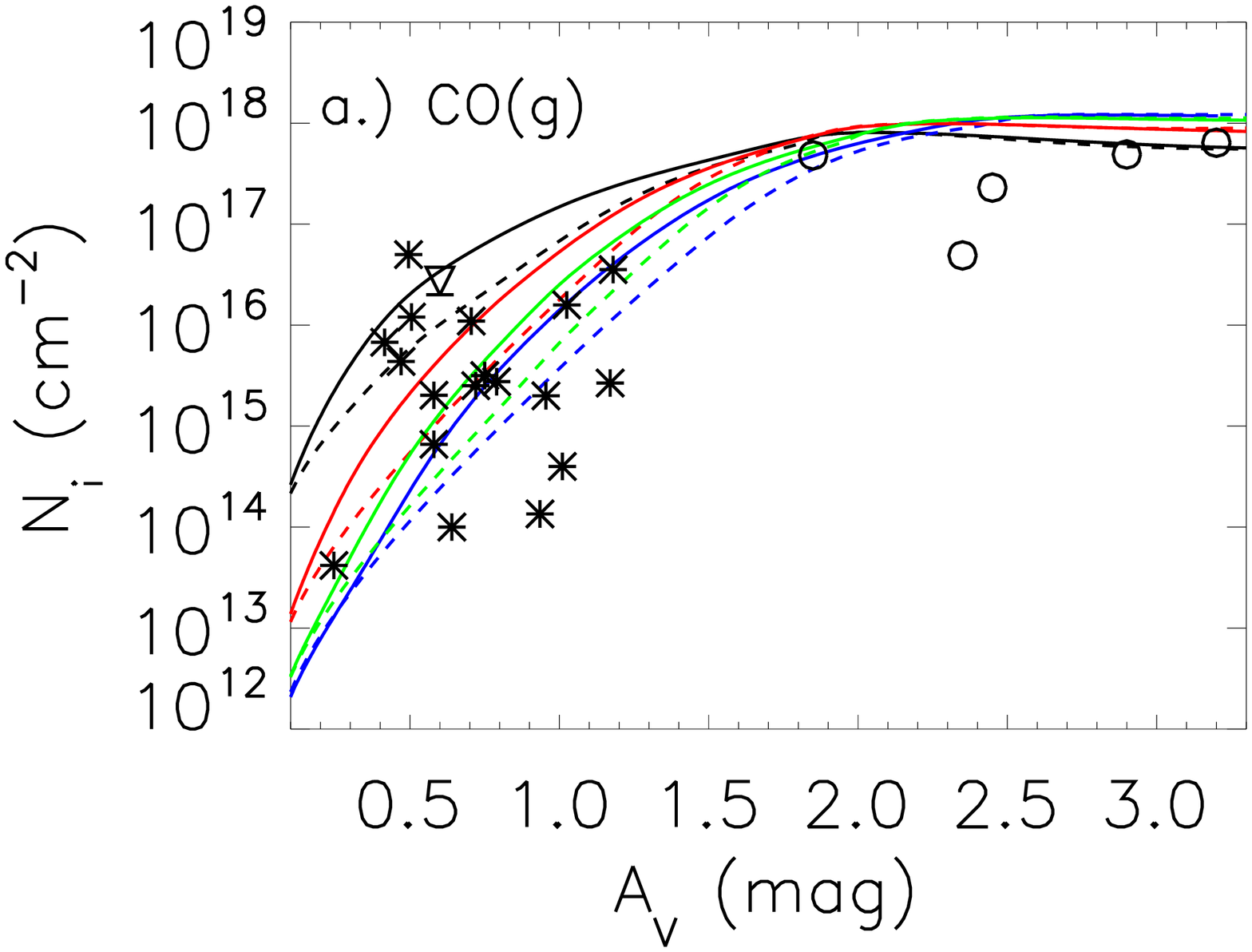}
\includegraphics[angle=90,scale=.4]{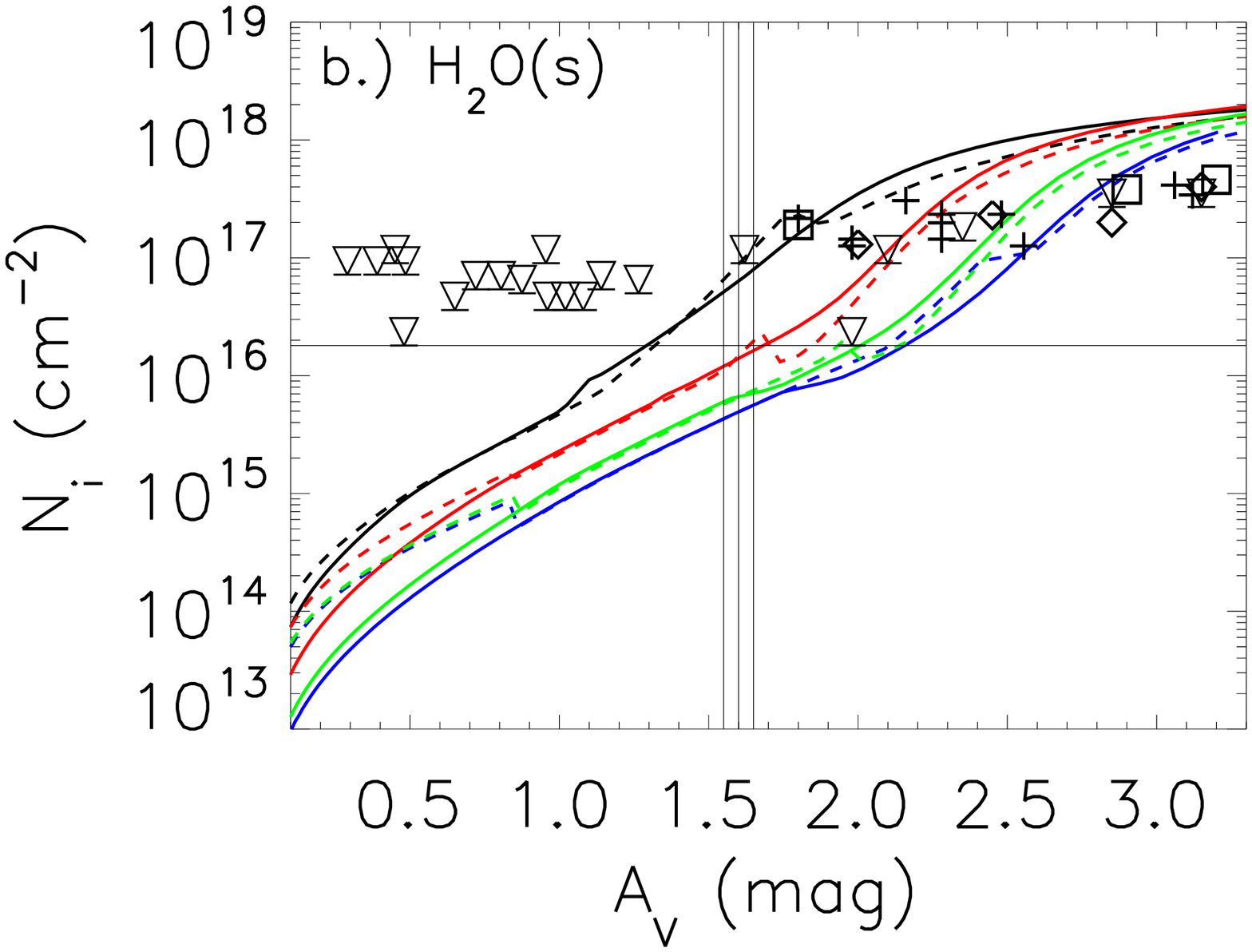}
\includegraphics[angle=90,scale=.4]{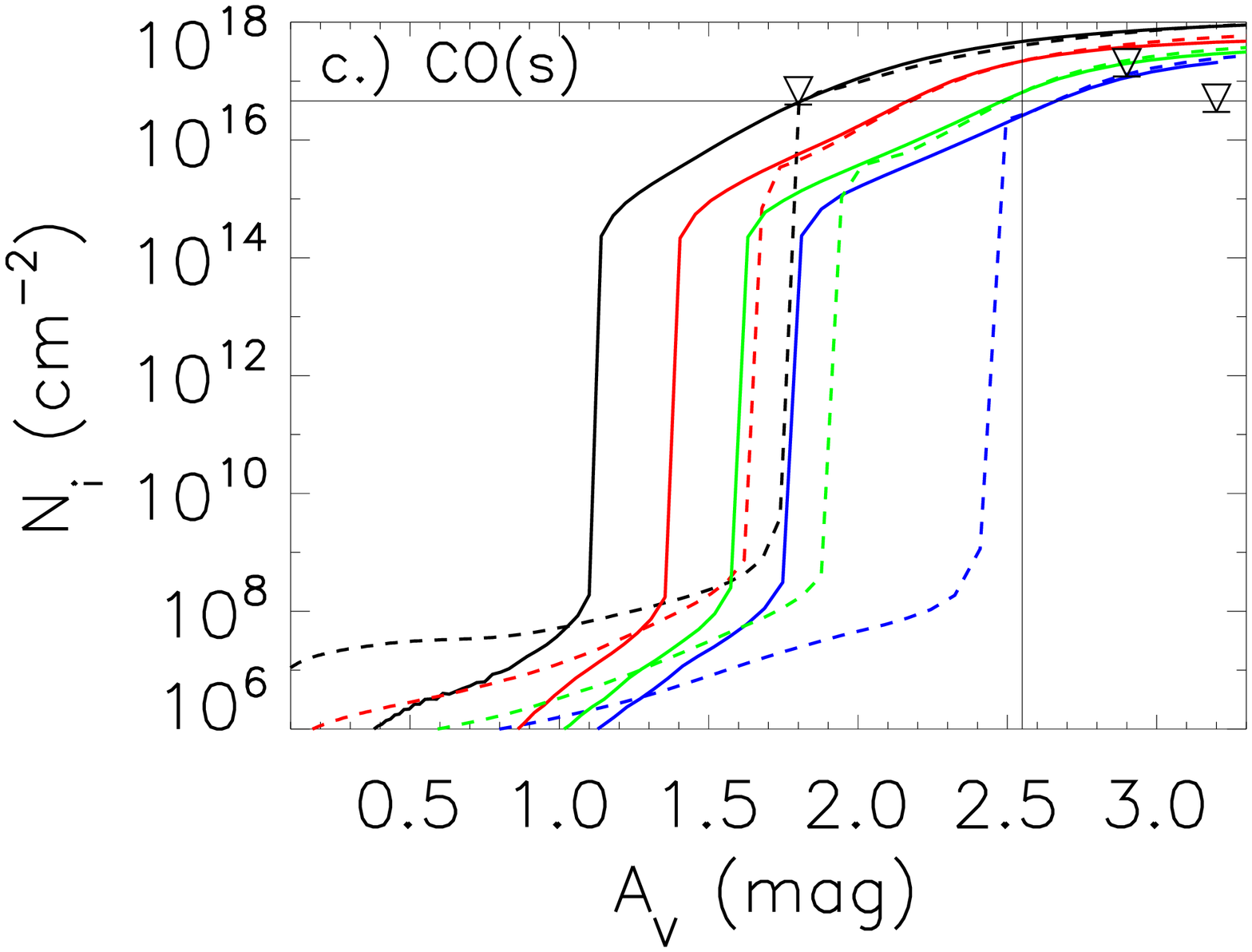}
\includegraphics[angle=90,scale=.4]{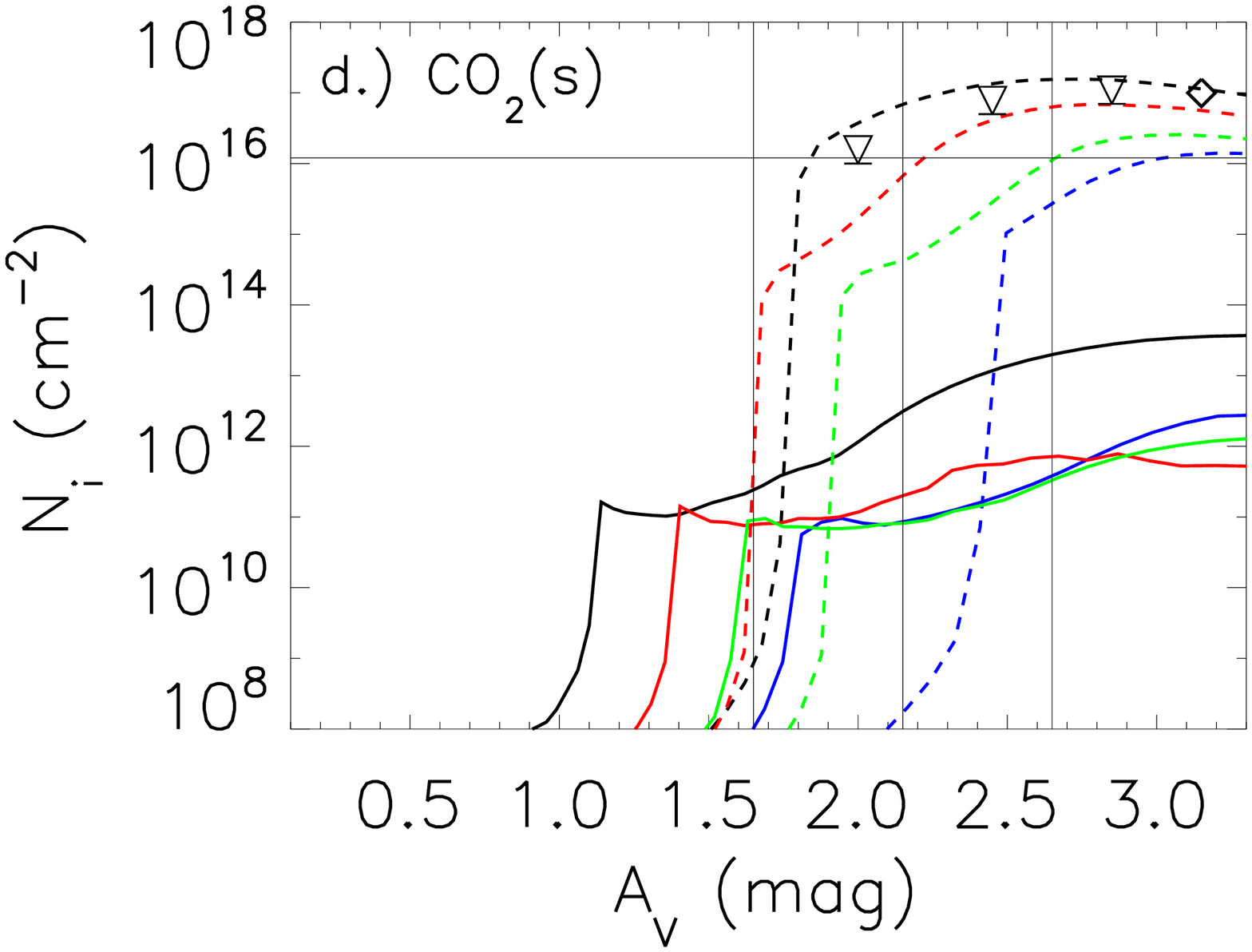}
\includegraphics[angle=90,scale=.4]{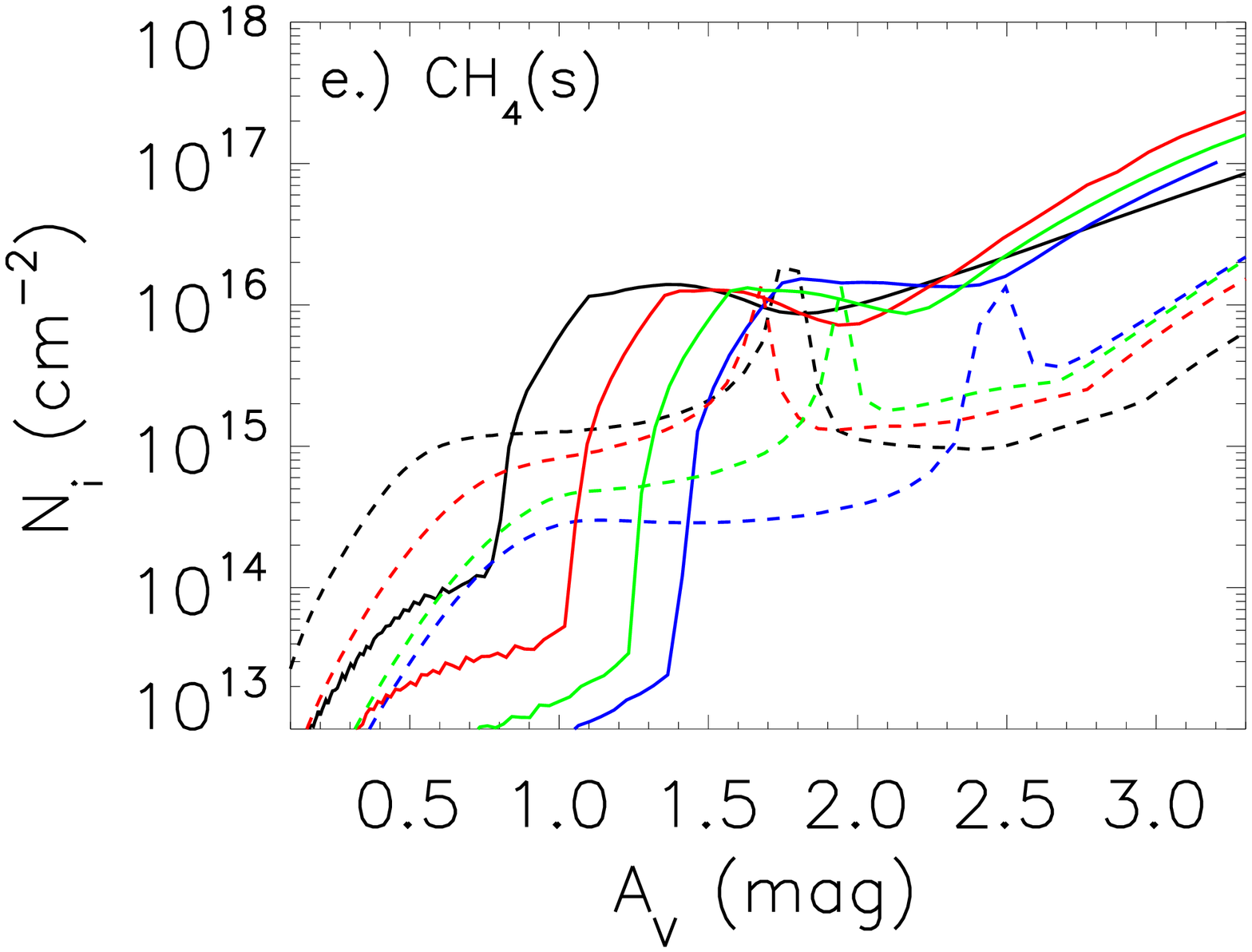}
\includegraphics[angle=90,scale=.4]{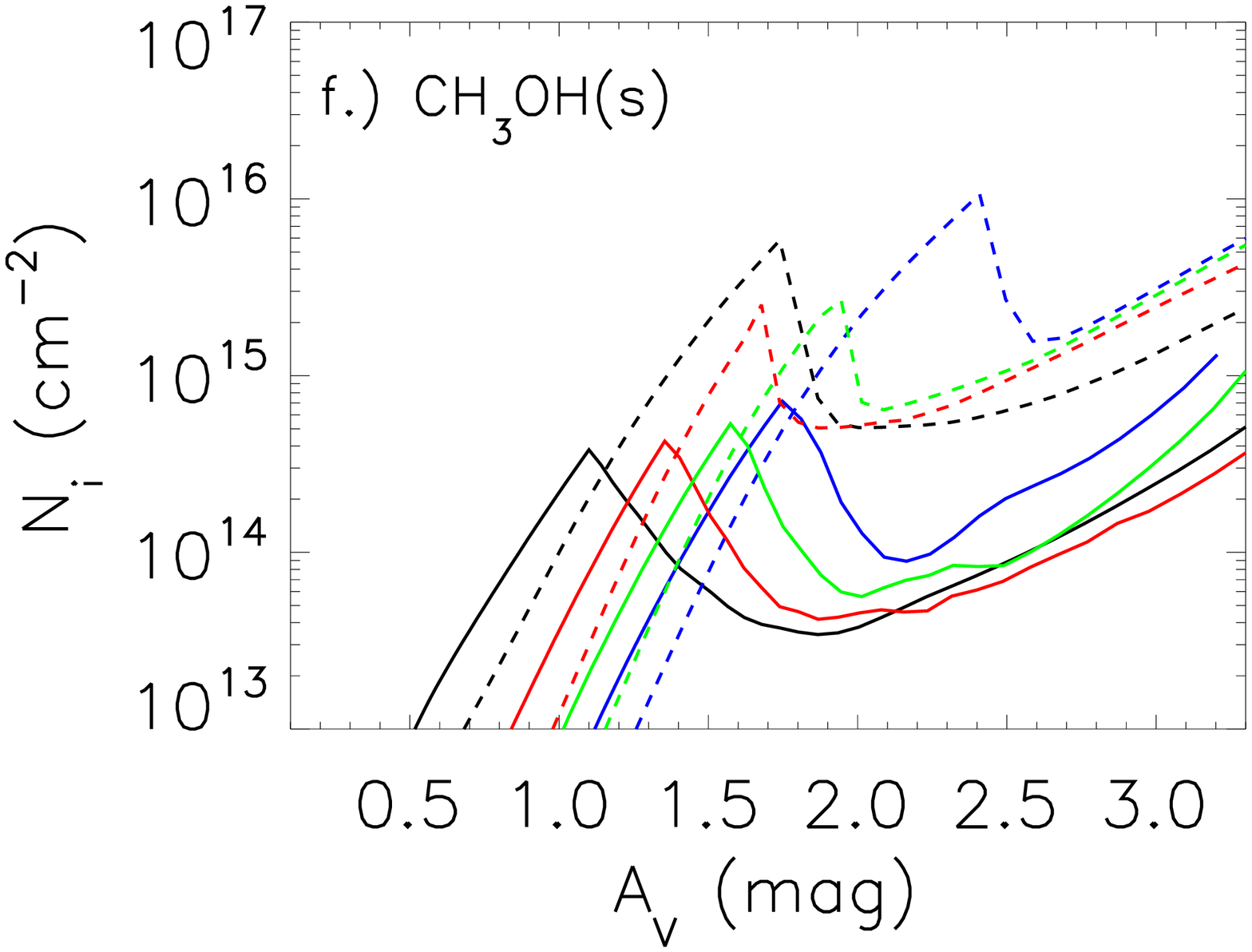}

\caption{
Column densities of selected molecules vs. visual extinction, for shock models 1 (green), 2 (red), 3 (blue), and 4 (black).  The solid lines represent models using the (A) parameters of $T_{\rm d,0} = 15$ K, $E_{\rm A} = 290$ K, while dashed lines represent models using the (B) parameters of $T_{\rm d,0} = 20$ K, $E_{\rm A} = 130$ K.  Observed ice values come from \citet{dougco207} (diamonds), \citet{Murakawa00} (crosses), and \citet{Teixeira99} (squares), while upper limits are included as inverted triangles. The observations of CO(g) come from \citet{Whittet89} (circles) and from a variety of translucent cloud sources (see Section~\ref{resultssectionobs} for citations.) Vertical lines denote the threshold extinction $A_{\rm th}$  \citep{dougco207} with estimated uncertainties; for CO(s) only the minimum value appears.  The horizontal lines indicate estimated minimum column densities for detection (see Section~\ref{resultssectionobs}).}
\label{fig-2NH}
\end{figure*}

\begin{figure*}
\centering
\includegraphics[angle=90,scale=.4]{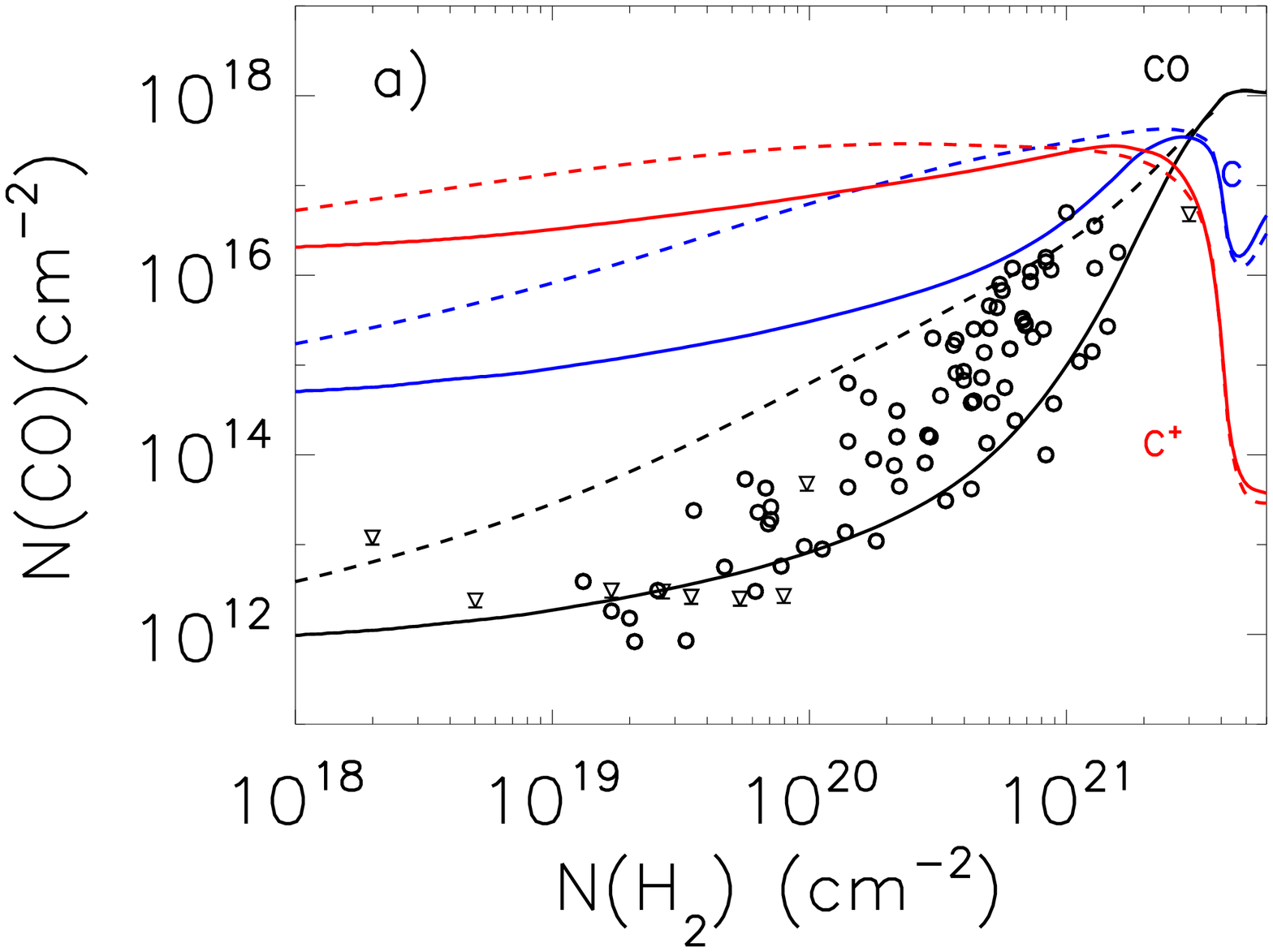}
\includegraphics[angle=90,scale=.4]{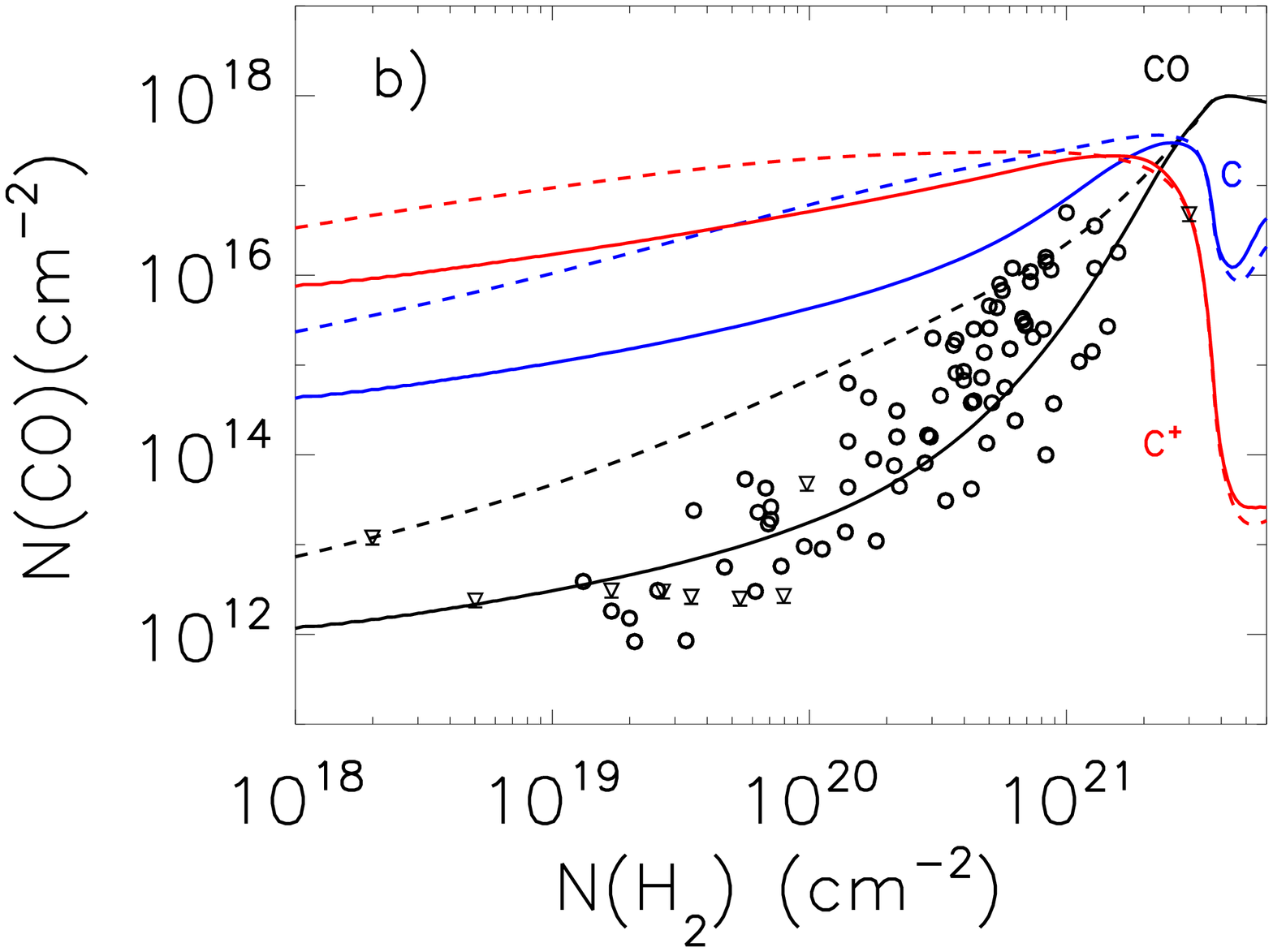}
\includegraphics[angle=90,scale=.4]{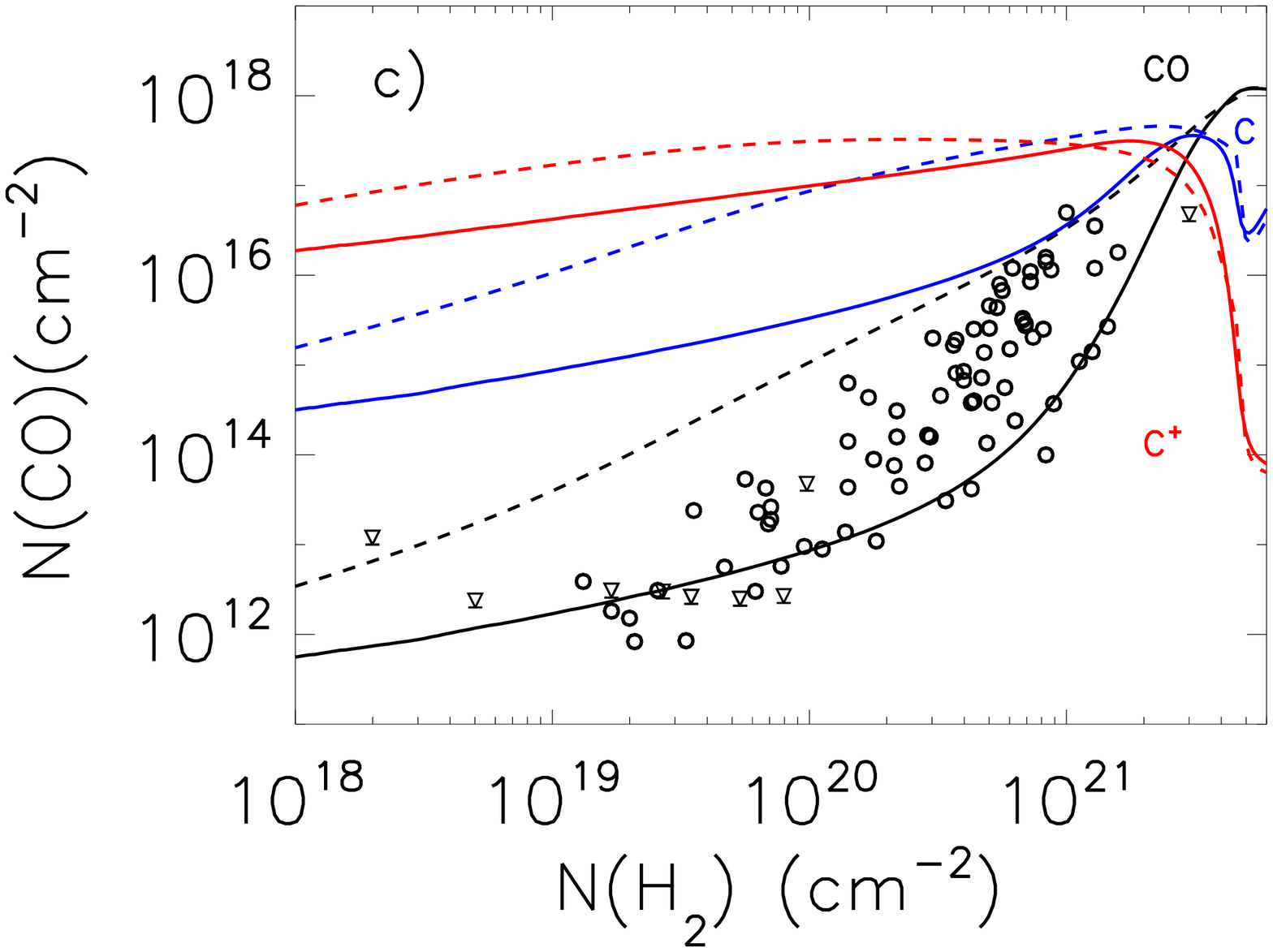}
\includegraphics[angle=90,scale=.4]{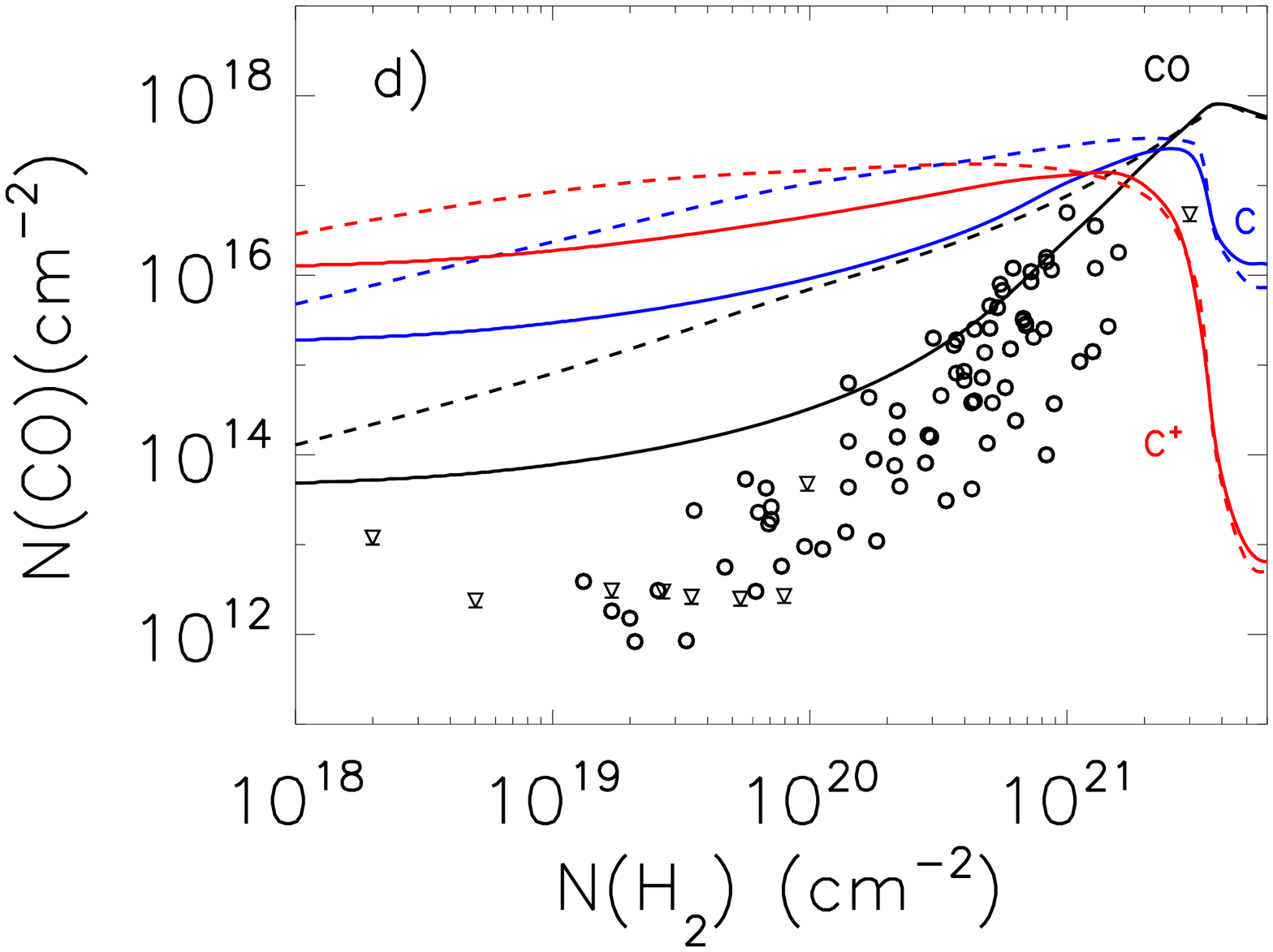}

\caption{
Column densities of CO(g) vs. H$_2$(g) for shock models 1 (panel a), 2 (panel b), 3 (panel c), and 4 (panel d).  The solid lines represent models using the (A) parameters of $T_{\rm d,0} = 15$ K, $E_{\rm A} = 290$ K, while dashed lines represent models using the (B) parameters of $T_{\rm d,0} = 20$ K, $E_{\rm A} = 130$ K.  Black lines represent CO(g), while the abundances of atomic C  and C$^+$  are represented as blue and red lines, respectively.  Observed values of CO(g) appear as circles while upper limits appear as inverted triangles; both are compiled from a variety of sources listed in Section~\ref{resultssectionobs}. }
\label{fig-3CO}
\end{figure*}

Our principal results are contained in two figures, in which we plot assorted column densities vs $A_V$ and $N[{\rm H_{2}}]$, respectively; the latter is more commonly used in studies of CO(g).  The column densities of CO(g), $N$[CO(g)], and an assortment of ices are plotted against $A_V$ in  Figure~\ref{fig-2NH} for the A and B parameter sets and all four shock models.  The column densities are shown for visual extinction through $A_{V} = 3.3$, as further extensions of the evolution are definitely beyond the scope of the shock hydrodynamics (see discussion in B04 regarding the importance of gravity).  The extinction used in Figure~\ref{fig-2NH} is considered to be ``edge-to-center'' for a line of sight into dense or translucent material.   Thus an observed extinction of 2 mag would correspond to 1 mag in the edge-to-center model frame.   The  column density for CO(g) is plotted  against $N$[H$_2$] in Figure~\ref{fig-3CO} along with column densities for the other major forms of carbon in the gas - C$^+$ and C. In both figures, solid lines represent models with the A parameter set, while dashed lines represent the B set.  

 In each figure, it can be seen that the difference in $N$[CO(g)] between the A and B parameter sets decreases as the extinction or H$_{2}$ column increases.  At low  H$_{2}$ column, the amount of CO(g) produced in the B models {\em apparently} exceeds that in the A models by up to 2 orders of magnitude. 
This effect occurs primarily because the warmer grains produce less H$_2$(g) at a given $A_V$ with our simple ``smooth'' grain model; there is much less of a temperature dependence in the ``rough'' grain model of  \citet{cch07}, which utilizes a microscopic Monte Carlo approach.   If the abscissa in Figure~\ref{fig-3CO} were the column of total hydrogen rather than of just H$_{2}$, the B and A model results for CO(g) would appear to be more similar.   Note also that the two plots with $N$[CO(g)] emphasize its growth with time in different ranges.  In particular, the range of labeled $A_{\rm V}$ values in  Figure~\ref{fig-2NH} of 0.5-3.0 corresponds to only a small portion of the H$_{2}$ column density range in Figure~\ref{fig-3CO}.    
  
  Despite these complications, a salient feature of  both figures is that CO(g) becomes the most abundant of the three major forms of carbon as $N[\rm{H_{2}}]\sim 4\times 10^{21}$ cm$^{-2}$ (corresponding to an edge-to-center $A_V \sim 2$).  Thus, long before a cold core with $A_{\rm V} \sim 10$ forms, the carbon inventory in the gas phase contains large amounts of CO(g).  Our results, obtained with a gas-grain model, are not identical with those of B04.  Nevertheless, they confirm how artificial the initial carbon inventory is for pseudo-time-dependent models, in which carbon is assumed to be totally in its atomic ionic form while the density and visual extinction are already at their final values.  Moreover, at a visual extinction
of 2, the [C$^+$ $+$ C]/[CO] abundance ratio is the same as we obtain at
steady-state with a gas-phase model at roughly the same density.  Thus, the
system has already evolved into a translucent cloud.

In addition to CO(g), calculated column densities for the major ice species H$_2$O(s), CO(s), and CO$_2$(s), along with the somewhat less abundant ices CH$_{4}$(s) and CH$_3$OH(s), are shown in Figure~\ref{fig-2NH}.  As illustrated in panel (b), the amount of H$_2$O(s)  grows relatively quickly to column densities in excess of 10$^{16}$ cm$^{-2}$ between $A_V = 1.3$ and 2.1 for the various models. As with CO(g), the results illustrate that this occurs most efficiently for shock models with greater total density.
The evolution of CO(s), as shown in panel (c),  tends to lag behind CO(g), and the plot of its column vs $A_{\rm V}$ shows a sharp threshold at intermediate visual extinction, which is correlated with a decrease in the rate of synthesis of methanol (see below).  The molecule forms more efficiently at lower extinction and earlier times for models with greater  density, similar to the trend seen for H$_2$O(s).  

There is a large distinction between models with the A and B parameters at low and intermediate visual extinction. 
For some molecular ices, specifically CO(s) and H$_2$O(s), as the visual extinction grows, the calculated column densities converge towards one another,
while others show order of magnitude agreement (CH$_4$, CH$_3$OH).   No convergence occurs  for 
CO$_2$(s), as shown in panel (d), which requires the adoption of the B parameter set to achieve a significant column density, mainly because the warmer grains allow faster diffusive reaction processes for heavy species.   Although the effects of individually varying the CO$_2$(s) formation barrier and turning off the Eley-Rideal mechanism on the abundance of CO$_2$(s) were considered, the grain temperature  is the most influential parameter for CO$_2$(s) formation.

  The formation of CO(g) and CO(s) also affects the evolution of  methane ice (CH$_4$(s)) and methanol ice (CH$_3$OH(s)),  shown in panels (e) and (f) of Figure~\ref{fig-2NH}, respectively.  Methane on grain surfaces is formed via sequential hydrogenation of C atoms, which accrete from the gas.  The less carbon and the more CO in the gas, the less efficient the surface formation of methane and the more CO(s) accretes onto grains.  Even at an extinction of 3.0, the CH$_{4}$(s) column densities for individual models tend to lie below  those of CO(s), especially for B parameter (warm grain) models.  As regards methanol, its only formation is via sequential hydrogenation of CO(s) by reactions involving H atoms on grain surfaces:
\begin{equation}
{\rm CO(s) \rightarrow HCO(s)  \rightarrow H_{2}CO(s)  \rightarrow H_{2}COH(s) \rightarrow CH_{3}OH(s) ,}
\end{equation}
so that the abundances of CO(s) and CH$_3$OH(s) should be intimately tied together.   
Nevertheless, the extinction dependence of the column densities of methane and methanol is far more complex than that of CO(s), especially with the B parameters, where extra peaks are seen.   For both CH$_4$(s) and CH$_3$OH(s) formation, warmer temperatures lead to more desorption of H atoms from the grains and would appear to slow down the hydrogenation of C(s) and CO(s).  This factor clearly affects the formation of methane at most times, but runs counter to the observation that higher surface temperatures aid methanol formation and CO(s) depletion.  Presumably this latter effect occurs because two of the reactions in the hydrogenation of methanol (H + CO and H + H$_2$CO) require overcoming a chemical reaction barrier.  

In addition to the results shown in Fig.~\ref{fig-2NH} and Fig.~\ref{fig-3CO}, we report additional results for major species at $A_{\rm V}$ = 3.  Specifically, Table~\ref{tbl-majform} contains the fractional abundances of the major gaseous and solid species for the A and B variants of Models 3 and 4.   These models span the range of physical conditions considered, and the table illustrates the differences in  composition that arise as a result.  This table provides a range of input abundances for subsequent dense core models.

\begin{table*}
\begin{center}
\caption{Fractional Abundances of Major Species at $A_V=3$ }             
\label{tbl-majform}      
\centering                        
\begin{tabular}{l c c c c }     
\hline\hline               
Species/Model & 3 A  &  3 B  &  4 A  &        4 B \\
\hline
Gas Phase \\
\hline  
H$_2$             & 0.5             & 0.5           & 0.5        & 0.5      \\
H                 & 7.2(-04)       & 7.4(-04)     & 1.1(-04)     & 1.1(-04)             \\
CO                & 1.1(-04)       & 1.1(-04)     & 5.1(-05)     & 4.8(-05)   \\
O                 & 1.0(-04)       & 1.0(-04)     & 1.2(-05)     & 1.3(-05)       \\
N$_2$             & 2.9(-05)       & 3.2(-05)     & 3.1(-05)     & 3.2(-05) \\
N                 & 2.5(-05)       & 2.1(-05)     & 4.0(-06)     & 4.5(-06)   \\
S                 & 8.7(-06)       & 8.9(-06)     & 6.0(-06)     & 6.2(-06)   \\
S$^+$             & 5.2(-06)       & 5.4(-06)     & 6.3(-07)     & 7.5(-07)\\
C                 & 3.4(-06)       & 2.5(-06)     & 1.2(-06)     & 6.3(-07) \\
CS                & 4.0(-08)       & 2.1(-08)     & 6.9(-08)     & 2.8(-08)  \\
CH$_4$            & 1.8(-08)       & 4.9(-09)     & 2.2(-08)     & 2.7(-09) \\
C$_2$H$_6$        & 8.9(-09)       & 5.0(-08)     & 4.0(-09)     & 2.4(-08) \\
C$^+$             & 8.8(-09)       & 6.7(-09)     & 5.6(-10)     & 4.3(-10)       \\
CO$_2$            & 2.2(-10)       & 1.1(-08)     & 1.7(-10)     & 4.1(-08)       \\
\hline
Solid Phase \\
\hline
H$_2$O(s)     & 4.7(-05)  & 4.2(-05)  & 1.4(-04) & 1.2(-04)  \\
CO(s)         & 9.9(-06)  & 1.1(-05)  & 6.8(-05) & 6.7(-05) \\
NH$_3$(s)     & 4.6(-06)  & 9.8(-07)  & 2.3(-06) & 6.9(-07)           \\
CH$_4$(s)     & 4.4(-06)  & 5.6(-07)  & 5.0(-06) & 2.7(-07)             \\
H$_2$S(s)     & 1.8(-06)  & 1.4(-06)  & 9.0(-06) & 8.7(-06)              \\
N$_2$(s)      & 1.3(-06)  & 1.7(-06)  & 9.3(-06) & 9.8(-06) \\
HCN(s)        & 1.0(-06)  & 1.2(-06)  & 3.7(-06) & 1.5(-06)  \\
MgH$_2$(s)    & 4.2(-07)  & 4.3(-07)  & 6.1(-07) & 6.2(-07) \\ 
SiH$_4$(s)    & 4.7(-08)  & 5.3(-08)  & 4.6(-07) & 4.6(-07)   \\
FeH(s)        & 4.4(-08)  & 4.3(-08)  & 8.9(-08) & 8.9(-08) \\
CH$_3$OH(s)   & 4.0(-08)  & 2.2(-07)  & 2.5(-08) & 1.3(-07)             \\
C$_2$H$_6$(s) & 2.8(-08)  & 1.3(-07)  & 1.7(-08)  & 9.4(-08)\\
H$_2$CO(s)    & 2.4(-08)  & 1.3(-07)  & 1.5(-08)  & 7.6(-08)\\
SiO(s)        & 2.2(-08)  & 2.6(-08)  & 2.4(-07) & 2.4(-07) \\
H$_2$CS(s)    & 2.0(-08)  & 5.2(-08)  & 1.7(-08)  & 9.3(-08) \\
CO$_2$(s)     & 9.7(-11)  & 7.8(-07)  & 2.9(-09) & 1.1(-05)\\
H$_2$O$_2$(s) & 7.8(-13)  & 8.0(-13)  & 8.8(-08) & 5.7(-07) \\  
\hline
\end{tabular}
\end{center}
\end{table*}

%\clearpage

%_____________________________________________________________

 Our results are obtained with a model that ignores grain growth.  But, as grains accumulate icy mantles, the cross sections grow likewise, potentially by substantial factors.  This effect is often omitted from gas-grain chemistry models.  In the single grain-size model considered in this paper, the cross section gradually doubles following $A_V \gtrsim 1.5$.   In some preliminary  simulations with mantle growth, we found that the onset of CO(s) and CO$_2$(s) growth occurs at slightly lower $A_V$, and that these ices reach abundances slightly different from those of Fig.~\ref{fig-2NH}, while CO(g) and water ice decrease.   Our overall  results and conclusions are not changed significantly by consideration of this effect.  The inclusion of mantle growth in a chemical model with a distribution of grain sizes is currently being undertaken by K. Acharyya and E. Herbst.  In addition to mantle growth, the effect of grain-grain collisions can lead to size distributions with considerably larger grains at time scales such as those considered here, according to detailed calculations by \cite{2009A&A...502..845O} (see, in particular, their Table 3).  Inclusion of this effect is beyond the scope of this paper.

\section{Comparison with Observations}
\label{resultssectionobs}

Although the density of the post-shock models can be as high as $2\times 10^{4}$ cm$^{-3}$, these do not simulate fully-formed dense cores given their low extinction, but rather the cloud  material such as that from which dense cores form.  Can relevant comparison be made with existing observational data, or are our results best used as initial conditions for models of normal dense cold cores?  The dark cloud in Taurus is perhaps the best region to explore this question for measurements of ices, while a large sample of CO(g) observations exists for diffuse and translucent clouds, as well as some lines of sight in Taurus.  One must also remember that our results represent cold material in the act of forming larger structures via a specific post-shock process, whereas there is no guarantee that the observed sources are transitional in this sense.  Indeed, at low visual extinction and/or H$_{2}$ column densities, our objects tend to be denser, smaller, and colder than well-studied diffuse clouds.

\subsection{CO(g) Observations}

 Estimates of $N$[CO(g)] along darker lines of sight in the Taurus dark cloud were presented by \citet{Whittet89} based on observations of \citet{Frerking82} and \citet{Crutcher85}.  Observations of $N$[CO(g)] in more translucent and diffuse clouds have been reported by \citet{Liszt08, Burgh07, Sheffer07, Sonnentrucker07, Gredel94}, and references therein.  The value of $A_V$ was reported for some of these sources by \citet{Rachford09}, while for the majority only $N$[H$_2$(g)] was reported.   

 The $N[\rm{CO(g)}]$ vs. $A_V$ data appear in Figure~\ref{fig-2NH}(a) as circles for the darker lines of sight, seen in the Taurus cloud, and asterisks for the translucent lines of sight except for one upper limit, shown as an inverted triangle.  It is apparent from this figure that many of the data points fall within the model curves in the lower $A_V$ range, and that all four shock  models with A and B parameter sets fit three of the darker line of sight observations to within a factor of 3 or better.

 The $N$[CO(g)] vs. $N$[H$_{2}$(g)] data appear in Figure~\ref{fig-3CO} as either circles or inverted triangles for upper limits.  The data, which are more numerous than those plotted against $A_{\rm V}$ and tend to contain lower CO(g) columns from diffuse and translucent cloud data, are fit best by model 2-A. The B-models tend to be somewhat worse than the A-models. Although the density, gas temperature, and size of the post-shock objects for the lower H$_{2}$ columns may not correspond with these parameters for actual observed clouds, the CO(g) columns we calculate are dependent mainly on the visual extinction, so observation and theory are in reasonable agreement anyway.  In addition, the calculated CO(g) columns  agree well with steady-state values for gas-phase models, if not quite as well as in Figure~\ref{fig-2NH}.

\subsection {Ice Observations}

Our predicted column densities for ices as functions of edge-to-center visual extinction can be compared with at least some infrared observational data along quiescent lines of sight in Taurus in our low visual extinction regime.  Shown in Figure~\ref{fig-2NH}, the limited data available in the extinction range up to 3.3 come from \citet{dougco207, Murakawa00, Teixeira99}, and references therein.  Some of these data are merely upper limits, while for CH$_{4}$(s) and CH$_{3}$OH(s), there are to the best of our knowledge not even upper limits in this range.    The  question arises as to whether observations at our relatively low extinction range belong to small cores or whether the material being sampled is simply the diffuse background.  That at least some of the Taurus observations at low extinction pertain to small cores has been shown by \citet{Whittet01, Whittet04}, who concluded that the lines of sight to HD 29647 and HD 283809 sample dense gas with extinctions of $A_V = 1.82$ and 2.85 (see their Figure 1 in the 2004 paper).  If representative of expanding objects as treated by our shock model, the size of the object with lower extinction, in the absence of self-gravity, would be in the range 0.05 pc (Model 4) to 0.5 pc (Model 3) (see Table~\ref{tbl-shockmods}). 
 
The vertical line or lines in panels (b)-(d) are empirically determined threshold extinctions, $A_{\rm th}$, and their uncertainty ranges for the Taurus dark cloud (the panel for CO(s) only shows the lower limit of this range).  The center lines represent the lowest $A_V$ values above which the specific ices are detectable in a cloud and are obtained by empirical fits to column density vs. visual extinction for a wide range of dark quiescent lines of sight in Taurus towards background stars at  larger visual extinction \citep{dougco207}.  
The specific edge-to-center thresholds for Taurus are $A_{\rm th}=1.6 \pm 0.05$ for H$_2$O(s), $3.4 \pm 0.8$ for CO(s), and $2.15 \pm 0.5$ mag for CO$_2$(s).  
These threshold values can  be regarded as conservative compared with some of the equivocal data at lower extinction from other groups \citep{Teixeira99, Murakawa00}, which might represent the diffuse background.  The thresholds are thought to roughly correspond with the accumulation of the equivalent of a few monolayers of a surface species, in a sudden onset above the threshold extinction.  To help interpret these thresholds, we have also plotted minimum detectable columns for the major ices, which are shown as horizontal lines. The estimates are made by substituting a minimal discernible optical depth of $\tau=0.01$ (Whittet, private communication)  into equation (5.5) of \citet{dustbook03}.   
At the time when $A_V \approx A_{\rm th}$, the horizontal lines in panels (b) and (d) represent the equivalent column density of about one or two monolayers of H$_2$O(s) and CO$_2$(s), while the line in panel (c) is closer to five layers of CO(s).

The observations of $N$[H$_2$O(s)] appear in Figure~\ref{fig-2NH}(b).  The Murakawa et al. detections were reported as optical depths, $\tau$, and we note that several of these values are less than the observational uncertainty of $\delta \tau=\pm 0.05$ reported. 
If we instead count these points as upper limits, as shown in the plot, the data below Whittet's threshold range are only limits, while firm detections appear at greater $A_{\rm V}$ than the threshold.  Below the threshold, the models results lie below these ``limits.''  Above the threshold,  the spread of data points and upper limits falls within the range of our model results, although it is difficult to determine which models are best.  
  Regarding the threshold itself, 
our calculated columns do not show a real threshold, but rise  gradually and reach the empirical threshold around the minimum detectable column (the horizontal line).  A proper interpretation may be that there is not really a threshold, so much as a drop below detectability masquerading as one.
Towards the highest extinction plotted, the assorted model results tend to converge to a narrow range of  values 3-10 $\times$ larger than the observed columns.   In some sense this is due to the chemistry readily hydrogenating oxygen atoms on grain surfaces.   At cold temperatures it is difficult to stop this process.  The difference between model and observations could be related to our assumed oxygen atom abundance being too high, perhaps through the presence of another reservoir of oxygen \citep[see, e.g.][]{dougco207}. 

. 
   
For the other major ices, CO(s) and CO$_{2}$(s), there are only upper limits in the extinction range $A_{\rm V}$ through $ \approx 3$, except for one detection of CO$_{2}$(s) toward Tamura 2 \citep{Whittet89, dougco207}.   The observational $N$[CO(s)] upper limits are not very constraining except perhaps at the highest visual extinction plotted where two upper limits are in better agreement with shock models 1 and 3.  For the case of CO$_2$(s), the B parameter models are required to bring the column density  to  near-detectable abundances.  The one  firmly detected observational data point is fit best by  models 4-B and 2-B.  
 
Figure~\ref{fig-2NH} shows very sharp thresholds for the onset of column densities of CO(s) and CO$_{2}$(s), but only for the latter are the calculated thresholds even in rough agreement with the empirical range of \citet{dougco207}.  For CO(s), we reach the minimum detectable column at visual extinction much smaller than the empirical threshold, which needs to be checked by data at lower extinction, where there are currently only  high upper limits to the CO(s) column. 
 It should be noted that, in the absence of photodesorption, minimum detectable columns for H$_{2}$O(s) and CO(s) are reached at lower visual extinction ($A_V \lesssim 1$), worsening the agreement with the threshold observations.

\section{Discussion}
\label{concsection}

The chemistry of cold dense cores has typically been treated by pseudo-time-dependent methods, in which physical conditions are homogeneous and time-independent.  Also, the initial chemical abundances are artificially assumed to take the form of atoms except for molecular hydrogen.  Although such gas-phase and gas-grain models of the chemistry are often in reasonable agreement with observations and even predictive in nature for the gas-phase species, the assumption of time-independent physical conditions ignores the formation of the core, a process that cannot be totally distinct from the chemistry that occurs.   
 For these reasons, we have started a program to determine how the chemistry is altered by consideration of the physical evolution of dense cores.
 
In this paper, we have reported our study of the chemistry that occurs during shock-induced formation of dark cloud material from the more diffuse background, with an emphasis on the composition of icy grain mantles and gaseous CO.  In this picture, gas-phase and grain-mantle molecular species form after the post-shock material cools and becomes dense, and as the visual extinction increases, in contrast with the assumption of constant physical conditions.  This stage can be considered as  preliminary  to the stage in which cold pre-stellar cores form.  Since our approach does not extend to the phase in which the self-gravity governs the evolution of these cooled objects, we cannot continue to increase their size and visual extinction beyond an extinction of $\approx 2-3$.   Although our calculated column densities cannot then be compared with those of ordinary cold dense, or prestellar, cores, we have compared them with the observed abundances of  diverse low-$A_V$ sources within the Taurus dark cloud, and a variety of other diffuse-to-translucent sources.

There are several key results of our calculations:
\begin{enumerate}
\item{Despite some divergence at lower extinction, all model results for CO(g) tend to converge as the edge-to-center extinction reaches $\sim 2.0$ (an H$_{2}$ column of $4 \times 10^{21}$ cm$^{-2}$) where CO(g)  becomes the dominant form of gaseous carbon.  We conclude that homogeneous gas-grain dense core models should not use atomic carbon as an initial abundance, but would be more realistic with initial abundances from the low-extinction results reported here. In particular, the calculated CO(g) columns and [C$^+$ + C]/CO ratios for the different models are not drastically different from the values obtained from gas-phase models at steady-state as a function of visual extinction or molecular hydrogen column density.
Our suggestion concerning the proper initial abundances to use has also appeared in a  recent paper by  \citet{2009A&A...508..783L}, who advocated that ``mechanisms for the chemical evolution from diffuse to dark gas should be included in model calculations'', a view that is the basis for our paper.} 

\item{By $A_V = 2.5$, all models show that  H$_2$O(s) and CO(s) have grown above their minimum observable abundances.   Models that form denser gas grow these ices at lower $A_V$.  These ices thus become abundant before  a full-fledged dense core can be produced.}
  
\item{The growth of CO(s) slows down the rate of synthesis of CH$_{4}$(s) because there is less of its precursor, neutral C, available.  By an extinction of 3.0, the methane column densities for individual models tend to lie below  those of CO(s), especially for the B parameter set, in which the dust temperature is higher.}

\item{The formation of CO$_2$(s) occurs efficiently only with the B parameter models,  especially with the higher density models, 2-B and 4-B.  The grain temperature is the most critical parameter for efficient CO$_2$(s) formation, with or without the Eley-Rideal mechanism.  }

\item{Models excluding photodesorption rates would form multiple layers of H$_2$O(s) and CO(s) at lower $A_V$ than either thresholds or the lowest reported firm detections \citep{dougco207}.  This result indicates that the inclusion of measured photodesorption rates is critical for realistic models.}

\end{enumerate}

Finally, and most importantly, our results show that if the shock model of B04 is correct, the early stages of both gas-phase and grain-surface chemistry occur as dense material is being formed, so that the chemistry of the cold cores that form from the final collapse of dense material should betray some indications of the physics of core formation.  Models that follow the chemistry as the dark cloud material  formed in our model collapses to produce  dense pre-stellar cores are desirable.
In addition, other dynamical scenarios than our shock model do exist, and we hope to explore them.

\begin{acknowledgements}
We thank the referee for some interesting points.  We are very thankful for helpful discussions with K. I. \"{O}berg and H. M. Cuppen regarding photodesorption experiments and D.C.B. Whittet and S.S. Shenoy regarding observational data.  We wish to thank P. Rimmer for assistance with the use of the Meudon PDR code.  We gratefully acknowledge the support of NASA through Spitzer data analysis contract 30331.  E.H. would like to thank the National Science Foundation for support of his research program in astrochemistry via grant  AST-0702876 and through the NSF Center for the Chemistry of the Universe (GA10750-131847).  
\end{acknowledgements}

%\begin{thebibliography}{}
%\bibliographystyle{aa}
%\bibliography{Hassel}

\begin{thebibliography}{51}
\expandafter\ifx\csname natexlab\endcsname\relax\def\natexlab#1{#1}\fi

\bibitem[{{Aikawa} {et~al.}(2005){Aikawa}, {Herbst}, {Roberts}, \&
  {Caselli}}]{Aikawa05}
{Aikawa}, Y., {Herbst}, E., {Roberts}, H., \& {Caselli}, P. 2005, \apj, 620,
  330

\bibitem[{{Andr{\'e}} {et~al.}(2008){Andr{\'e}}, {Basu}, \& {Inutsuka}}]{ABI09}
{Andr{\'e}}, P., {Basu}, S., \& {Inutsuka}, S.~I. 2008, arXiv:(0801.4210)

\bibitem[{{Bergin} {et~al.}(2004){Bergin}, {Hartmann}, {Raymond}, \&
  {Ballesteros-Paredes}}]{Bergin04}
{Bergin}, E.~A., {Hartmann}, L.~W., {Raymond}, J.~C., \& {Ballesteros-Paredes},
  J. 2004, \apj, 612, 921 (B04)

\bibitem[{{Bergin} \& {Tafalla}(2007)}]{BT07}
{Bergin}, E.~A. \& {Tafalla}, M. 2007, \araa, 45, 339

\bibitem[{{Bohren} \& {Huffman}(1983)}]{bohhuf83}
{Bohren}, C.~F. \& {Huffman}, D.~R. 1983, Absorption and scattering of light by
  small particles (New York: Wiley)

\bibitem[{{Burgh} {et~al.}(2007){Burgh}, {France}, \& {McCandliss}}]{Burgh07}
{Burgh}, E.~B., {France}, K., \& {McCandliss}, S.~R. 2007, \apj, 658, 446

\bibitem[{{Chang} {et~al.}(2007){Chang}, {Cuppen}, \& {Herbst}}]{cch07}
{Chang}, Q., {Cuppen}, H.~M., \& {Herbst}, E. 2007, \aap, 469, 973

\bibitem[{{Crutcher}(1985)}]{Crutcher85}
{Crutcher}, R.~M. 1985, \apj, 288, 604

\bibitem[{{Cuppen} {et~al.}(2006){Cuppen}, {Morata}, \& {Herbst}}]{cuppen06}
{Cuppen}, H.~M., {Morata}, O., \& {Herbst}, E. 2006, \mnras, 367, 1757

\bibitem[{{Draine}(1978)}]{draine78}
{Draine}, B.~T. 1978, \apjs, 36, 595

\bibitem[{{Draine} \& {Bertoldi}(1996)}]{drainebertoldi96}
{Draine}, B.~T. \& {Bertoldi}, F. 1996, \apj, 468, 269

\bibitem[{{Draine} \& {Lee}(1984)}]{drainelee84}
{Draine}, B.~T. \& {Lee}, H.~M. 1984, \apj, 285, 89

\bibitem[{{Fournier} {et~al.}(1979){Fournier}, {Deson}, {Vermeil}, \&
  {Pimentel}}]{FEA79}
{Fournier}, J., {Deson}, J., {Vermeil}, C., \& {Pimentel}, G.~C. 1979,
  J.Chem.Phys., 70, 5726

\bibitem[{{Frerking} {et~al.}(1982){Frerking}, {Langer}, \&
  {Wilson}}]{Frerking82}
{Frerking}, M.~A., {Langer}, W.~D., \& {Wilson}, R.~W. 1982, \apj, 262, 590

\bibitem[{{Garrod} \& {Herbst}(2006)}]{GH06}
{Garrod}, R.~T. \& {Herbst}, E. 2006, \aap, 457, 927

\bibitem[{{Garrod} {et~al.}(2006){Garrod}, {Park}, {Caselli}, \&
  {Herbst}}]{gpch06}
{Garrod}, R.~T., {Park}, I.~H., {Caselli}, P., \& {Herbst}, E. 2006, Faraday
  Discuss., 133, 51

\bibitem[{{Garrod} {et~al.}(2007){Garrod}, {Wakelam}, \& {Herbst}}]{gwh07}
{Garrod}, R.~T., {Wakelam}, V., \& {Herbst}, E. 2007, \aap, 467, 1103

\bibitem[{{Gredel} {et~al.}(1994){Gredel}, {van Dishoeck}, \&
  {Black}}]{Gredel94}
{Gredel}, R., {van Dishoeck}, E.~F., \& {Black}, J.~H. 1994, \aap, 285, 300

\bibitem[{{Grim} \& {D'Hendecourt}(1986)}]{gdh86}
{Grim}, R.~J.~A. \& {D'Hendecourt}, L.~B. 1986, \aap, 167, 161

\bibitem[{{Hartmann} {et~al.}(2001){Hartmann}, {Ballesteros-Paredes}, \&
  {Bergin}}]{hbb01}
{Hartmann}, L., {Ballesteros-Paredes}, J., \& {Bergin}, E.~A. 2001, \apj, 562,
  852

\bibitem[{{Hasegawa} {et~al.}(1992){Hasegawa}, {Herbst}, \& {Leung}}]{hhl92}
{Hasegawa}, T.~I., {Herbst}, E., \& {Leung}, C.~M. 1992, \apjs, 82, 167

\bibitem[{{Le Petit} {et~al.}(2006){Le Petit}, {Nehm{\'e}}, {Le Bourlot}, \&
  {Roueff}}]{Meudon}
{Le Petit}, F., {Nehm{\'e}}, C., {Le Bourlot}, J., \& {Roueff}, E. 2006, \apjs,
  164, 506

\bibitem[{{Lee} {et~al.}(1996){Lee}, {Herbst}, {Pineau des Forets}, {Roueff},
  \& {Le Bourlot}}]{lee96}
{Lee}, H., {Herbst}, E., {Pineau des Forets}, G., {Roueff}, E., \& {Le
  Bourlot}, J. 1996, \aap, 311, 690

\bibitem[{{Liszt}(2008)}]{Liszt08}
{Liszt}, H.~S. 2008, \aap, 492, 743

\bibitem[{{Liszt}(2009)}]{2009A&A...508..783L}
{Liszt}, H.~S. 2009, \aap, 508, 783

\bibitem[{{McKee} \& {Ostriker}(2007)}]{MO07}
{McKee}, C.~F. \& {Ostriker}, E.~C. 2007, \araa, 45, 565

\bibitem[{{Mitchell} \& {Watt}(1985)}]{MitchellWatt}
{Mitchell}, G.~F. \& {Watt}, G.~D. 1985, \aap, 151, 121

\bibitem[{{Murakawa} {et~al.}(2000){Murakawa}, {Tamura}, \&
  {Nagata}}]{Murakawa00}
{Murakawa}, K., {Tamura}, M., \& {Nagata}, T. 2000, \apjs, 128, 603

\bibitem[{{Nejad} \& {Williams}(1992)}]{NW92}
{Nejad}, L.~A.~M. \& {Williams}, D.~A. 1992, \mnras, 255, 441

\bibitem[{{Nguyen} {et~al.}(2002){Nguyen}, {Ruffle}, {Herbst}, \&
  {Williams}}]{Nguyen02}
{Nguyen}, T.~K., {Ruffle}, D.~P., {Herbst}, E., \& {Williams}, D.~A. 2002,
  \mnras, 329, 301

\bibitem[{{Nummelin} {et~al.}(2001){Nummelin}, {Whittet}, {Gibb}, {Gerakines},
  \& {Chiar}}]{Nummelin01}
{Nummelin}, A., {Whittet}, D.~C.~B., {Gibb}, E.~L., {Gerakines}, P.~A., \&
  {Chiar}, J.~E. 2001, \apj, 558, 185

\bibitem[{{{\"O}berg} {et~al.}(2007){{\"O}berg}, {Fuchs}, {Awad}, {Fraser},
  {Schlemmer}, {van Dishoeck}, \& {Linnartz}}]{ObergCO07}
{{\"O}berg}, K.~I., {Fuchs}, G.~W., {Awad}, Z., {et~al.} 2007, \apjl, 662, L23

\bibitem[{{{\"O}berg} {et~al.}(2009{\natexlab{a}}){{\"O}berg}, {Linnartz},
  {Visser}, \& {van Dishoeck}}]{ObergWater}
{{\"O}berg}, K.~I., {Linnartz}, H., {Visser}, R., \& {van Dishoeck}, E.~F.
  2009{\natexlab{a}}, \apj, 693, 1209

\bibitem[{{{\"O}berg} {et~al.}(2009{\natexlab{b}}){{\"O}berg}, {van Dishoeck},
  \& {Linnartz}}]{ObergPHO}
{{\"O}berg}, K.~I., {van Dishoeck}, E.~F., \& {Linnartz}, H.
  2009{\natexlab{b}}, \aap, 496, 281

\bibitem[{{Ormel} {et~al.}(2009){Ormel}, {Paszun}, {Dominik}, \&
  {Tielens}}]{2009A&A...502..845O}
{Ormel}, C.~W., {Paszun}, D., {Dominik}, C., \& {Tielens}, A.~G.~G.~M. 2009,
  \aap, 502, 845

\bibitem[{{Pineau des For\^{e}ts} {et~al.}(1991){Pineau des For\^{e}ts},
  {Flower}, \& {Herbst}}]{Pineau91}
{Pineau des For\^{e}ts}, G., {Flower}, D.~R., \& {Herbst}, E. 1991, \mnras,
  253, 359

\bibitem[{{Rachford} {et~al.}(2009){Rachford}, {Snow}, {Destree}, {Ross},
  {Ferlet}, {Friedman}, {Gry}, {Jenkins}, {Morton}, {Savage}, {Shull},
  {Sonnentrucker}, {Tumlinson}, {Vidal-Madjar}, {Welty}, \&
  {York}}]{Rachford09}
{Rachford}, B.~L., {Snow}, T.~P., {Destree}, J.~D., {et~al.} 2009, \apjs, 180,
  125

\bibitem[{{Reach} {et~al.}(1995){Reach}, {Dwek}, {Fixsen}, {Hewagama},
  {Mather}, {Shafer}, {Banday}, {Bennett}, {Cheng}, {Eplee}, {Leisawitz},
  {Lubin}, {Read}, {Rosen}, {Shuman}, {Smoot}, {Sodroski}, \&
  {Wright}}]{Reach95}
{Reach}, W.~T., {Dwek}, E., {Fixsen}, D.~J., {et~al.} 1995, \apj, 451, 188

\bibitem[{{Roser} {et~al.}(2001){Roser}, {Vidali}, {Manic{\`o}}, \&
  {Pirronello}}]{RVMP01}
{Roser}, J.~E., {Vidali}, G., {Manic{\`o}}, G., \& {Pirronello}, V. 2001,
  \apjl, 555, L61

\bibitem[{{Ruffle} \& {Herbst}(2001)}]{RH01III}
{Ruffle}, D.~P. \& {Herbst}, E. 2001, \mnras, 324, 1054

\bibitem[{{Sheffer} {et~al.}(2007){Sheffer}, {Rogers}, {Federman}, {Lambert},
  \& {Gredel}}]{Sheffer07}
{Sheffer}, Y., {Rogers}, M., {Federman}, S.~R., {Lambert}, D.~L., \& {Gredel},
  R. 2007, \apj, 667, 1002

\bibitem[{{Sonnentrucker} {et~al.}(2007){Sonnentrucker}, {Welty}, {Thorburn},
  \& {York}}]{Sonnentrucker07}
{Sonnentrucker}, P., {Welty}, D.~E., {Thorburn}, J.~A., \& {York}, D.~G. 2007,
  \apjs, 168, 58

\bibitem[{{Teixeira} \& {Emerson}(1999)}]{Teixeira99}
{Teixeira}, T.~C. \& {Emerson}, J.~P. 1999, \aap, 351, 292

\bibitem[{{Tielens} \& {Hagen}(1982)}]{TH82}
{Tielens}, A.~G.~G.~M. \& {Hagen}, W. 1982, \aap, 114, 245

\bibitem[{{Whittet}(2003)}]{dustbook03}
{Whittet}, D.~C.~B. 2003, Dust in the galactic environment (Bristol: Institute
  of Physics Publishing)

\bibitem[{{Whittet} {et~al.}(1989){Whittet}, {Adamson}, {Duley}, {Geballe}, \&
  {McFadzean}}]{Whittet89}
{Whittet}, D.~C.~B., {Adamson}, A.~J., {Duley}, W.~W., {Geballe}, T.~R., \&
  {McFadzean}, A.~D. 1989, \mnras, 241, 707

\bibitem[{{Whittet} {et~al.}(2001){Whittet}, {Gerakines}, {Hough}, \&
  {Shenoy}}]{Whittet01}
{Whittet}, D.~C.~B., {Gerakines}, P.~A., {Hough}, J.~H., \& {Shenoy}, S.~S.
  2001, \apj, 547, 872

\bibitem[{{Whittet} {et~al.}(2007){Whittet}, {Shenoy}, {Bergin}, {Chiar},
  {Gerakines}, {Gibb}, {Melnick}, \& {Neufeld}}]{dougco207}
{Whittet}, D.~C.~B., {Shenoy}, S.~S., {Bergin}, E.~A., {et~al.} 2007, \apj,
  655, 332

\bibitem[{{Whittet} {et~al.}(2004){Whittet}, {Shenoy}, {Clayton}, \&
  {Gordon}}]{Whittet04}
{Whittet}, D.~C.~B., {Shenoy}, S.~S., {Clayton}, G.~C., \& {Gordon}, K.~D.
  2004, \apj, 602, 291

\bibitem[{{Williams}(1988)}]{Wi88}
{Williams}, D.~A. 1988, in Rate Coefficients in Astrochemistry, ed. T.~J.
  {Millar} \& D.~A. {Williams} (Dordrecht: Kluwer), 281

\bibitem[{{Zucconi} {et~al.}(2001){Zucconi}, {Walmsley}, \&
  {Galli}}]{Zucconi01}
{Zucconi}, A., {Walmsley}, C.~M., \& {Galli}, D. 2001, \aap, 376, 650

\end{thebibliography}

%\end{thebibliography}

\end{document}